%% file: main.tex
\documentclass[sigconf]{acmart}

\usepackage[utf8]{inputenc}
\usepackage{hyperref}

\AtBeginDocument{%
  \providecommand\BibTeX{{%
    \normalfont B\kern-0.5em{\scshape i\kern-0.25em b}\kern-0.8em\TeX}}}

\copyrightyear{2022}
\acmYear{2022}
\setcopyright{rightsretained}
\acmConference[CHI '22]{CHI Conference on Human Factors in Computing Systems}{April 29-May 5, 2022}{New Orleans, LA, USA}
\acmBooktitle{CHI Conference on Human Factors in Computing Systems (CHI '22), April 29-May 5, 2022, New Orleans, LA, USA}
\acmDOI{10.1145/3491102.3501857}
\acmISBN{978-1-4503-9157-3/22/04}

\begin{document}

\title{(Re)Politicizing Digital Well-Being: Beyond User Engagements}


\author{Niall Docherty}
\affiliation{%
  \institution{Microsoft Research}
  \city{Cambridge}
  \state{Massachusetts}
  \country{USA}
}
\email{ndocherty@microsoft.com}

\author{Asia J. Biega}
\affiliation{%
  \institution{Max Planck Institute for Security and Privacy}
  \city{Bochum}
  \country{Germany}
  }
\email{asia.biega@mpi-sp.org}



\begin{abstract}
The psychological costs of the attention economy are often considered through the binary of harmful design and healthy use, with digital well-being chiefly characterised as a matter of personal responsibility. This article adopts an interdisciplinary approach to highlight the empirical, ideological, and political limits of embedding this individualised perspective in computational discourses and designs of digital well-being measurement. We will reveal well-being to be a culturally specific and environmentally conditioned concept and will problematize user engagement as a universal proxy for well-being. Instead, the contributing factors of user well-being will be located in environing social, cultural, and political conditions far beyond the control of individual users alone. In doing so, we hope to reinvigorate the issue of digital well-being measurement as a nexus point of political concern, through which multiple disciplines can study experiences of digital ill as symptomatic of wider social inequalities and (capitalist) relations of power.
\end{abstract}






\maketitle

\pagestyle{plain}

\input{contents/00_introduction}
\input{contents/05_well-being}
\input{contents/10_engagement}
\input{contents/15_discussion}

\input{contents/20_conclusions}

\bibliographystyle{ACM-Reference-Format}
\bibliography{main}

\end{document}

%% file: contents/00_introduction.tex
\section{Introduction}
\label{sec:introduction}

Concerns surrounding digital well-being have emerged in recent years to signal that something may be amiss in contemporary relationships with technology~\cite{humane-tech-center}. Critics of technological addiction target haptically entrapping computational designs as significant factors in the perceived decline in user well-being, while capitalist technology companies and platforms argue that it is up to users, and users alone, to take personal responsibility for their digital health through self-controlled engagement~\cite{docherty_facebooks_2020}. This stance, which prioritizes a notion of \textit{time well spent }– now a key phrase used by leaders of technology companies such as Mark Zuckerberg~\cite{mosseri_bringing_2018}, absolves platforms of any serious accountability for the well-being of their users, while simultaneously ensuring flows of profitable active use remain open. In this way, discourses of temporal self-control serve a strategic economic function. Strangely perhaps, this type of self-controlled engagement is endorsed by even the critics of the so-called attention economy, who express concern over the potential well-being harms that come with the datafication of everyday life ~\cite{docherty_digital_2021}. Because of this dual discursive function, from both sides of the debate, `taking back control' of personal platform use exists as the chief safeguard against psychological digital damage in the present age. In response, recent well-being interventions from the HCI community have developed frameworks for pre-emptive positive computing~\cite{calvo_positive_2014}, time-management controls, user well-being dashboards and engagement analytics, and reducing potential and actual user harms~\cite{stephanidis_seven_2019}. As such, individual digital well-being is often centered around the question of whether or not users are spending their time on platforms in a controlled, positive manner.

This approach is reflected in recent governmental policy initiatives aimed at reducing pathological engagement in the US~\cite{us-social-media-addiction}, UK~\cite{uk-immersive-report}, and the EU~\cite{eu-ai-act}. At the same time, non-governmental organizations, such as the Center for Humane Technology~\cite{humane-tech-center}, lobby for greater awareness of the pitfalls of unconscious, passive, engagement with technology. Determining the quality of user time and engagements on platforms, accordingly, resonates with digital addiction frameworks~\cite{alrobai2014digital}. The Internet Addiction Questionnaire, for example, measures different dimensions of time spent on a platform, whether platform usage influences a user's life offline, or whether a user thinks about a platform while not using it~\cite{young2009internet}. Other studies adopt similar temporal, disconnection, and engagement-based measures of well-being and addiction~\cite{allcott2020welfare,mok2021complementary}, with more moralized discourses, such as the aforementioned time well spent movement, calling for personal reductions of certain negative types of user engagement.

Much like reasoning about fairness has inaugurated the development of fairness metrics and algorithmic interventions, the recent focus on user engagements responds to the imagined need to find ways to quantify formal dimensions of user well-being. If this path is followed, a pertinent question is \emph{whether user engagement is an adequate and sufficient proxy for well-being?} By contrasting two established measurement models---of digital addiction and user engagement---this paper argues that answering this question is not as straightforward as it may seem. Primarily, this is because precisely what we are measuring when we seek to measure well-being is unclear. Well-being, initially appearing to be a common sense concept, is, in fact, a highly contested, unstable term~\cite{dodge_challenge_2012}. How one treats the concept largely depends on the field of its application.~\cite{alexandrova_well-being_2012} For instance, public health scholars explore the political ramifications of adopting one view of well-being over another in policy decisions~\cite{fisher_theory_2019}, psychology seeks to categorize and measure personal experiences of flourishing for different groups ~\cite{ryff_know_2008}, while philosophy explores what it means to live well as a human being in time ~\cite{fletcher_routledge_2015}. 

Computer science offers the term ‘digital well-being’, and specifically refers ``to the impact of digital technologies on what it means to live a life that is good for a human being'' (\citet{burr_ethics_2020} p. 2313). However, rather than take this at face value, this paper argues that we should pay serious attention to the fact that the human good is a relative concept that cannot be divorced from the time and place of its articulation. Configuring well-being in computational systems, as such, always entails normative determinations that have significant impacts for users. Accordingly, this paper argues that practitioners ought to make the normative stakes and ethical implications of digital well-being explicit at every stage of the design process. In making this case, we will show that problematizing digital well-being opens up a whole range of related questions surrounding what it means to live well with technology in current sociotechnical networks. We suggest that exploring these questions is a key research agenda that demands further exploration

We develop an interdisciplinary approach in order to raise awareness of the operative social, political, and philosophical issues that accompany the inherent normative function of discourses and designs of digital well-being. To do this, we specifically draw upon modes of analysis associated with the related disciplines of Critical Theory and Cultural Studies. Social science disciplines have employed digital processes as both objects and tools of study~\cite{lupton_digital_2015}, and there exists numerous instances of sociological methods informing system design~\cite{sommerville_sociologists_1993} and in the implementation of socially conscious digital infrastructures~\cite{goulden_wild_2017}. In an Information System (IS) context, ~\citet{richardson_introduction_2006} highlight three key benefits that combining Critical Theory with Computer Science can offer: insight, critique and transformative redefinition. They write:

\begin{quote}
``Insight helps to highlight hidden or less obvious aspects of social reality in the process of seeing how various forms of knowledge, objects, and events are formed and sustained. Critique challenges many of the taken-for-granted assumptions, beliefs, ideologies, discourses that permeate IS phenomena. Transformative redefinition is the development of critical, relevant knowledge and practical understanding to facilitate emancipatory change'' (\cite{richardson_introduction_2006} p.4) 
\end{quote}
 Engaging with this interdisciplinary approach, this article will critically examine the current focus on user engagement as a proxy for well-being within technological circles, and question whether targeting modes of user engagement is a useful framework to understand, and potentially improve, user well-being on platforms. In looking beyond well-being as a stable, universally accepted, category of human wellness, we hope to help ground future attempts to design for user well-being in a critical praxis.

The opening sections will provide a concise overview of the definitional controversy of well-being, drawing upon philosophy, psychology, and the history of ideas to show how competing conceptualizations of well-being can broadly be categorized into eudaimonic, hedonistic, and structural accounts. Outlining the characteristics of these three accounts, exploring their philosophical roots, and detailing their respective modes of measurement will demonstrate that the notion of well-being should by no means be taken-for-granted in HCI. Rather, the specific vision of human well-being one adopts has significant implications for the design of computational systems that are built with its amelioration in mind. The article will then explore the notion of digital well-being in more depth and move on to examine the feasibility of employing user engagement metrics as proxies for user well-being through the clinical lens of digital addiction. The limits of a purely behaviorist account of digital well-being will be highlighted in the discussion section, where we will locate the determinants of well-being in environing social, cultural, and political conditions far beyond the control, or even behavioral choices, of individual users alone. We will close by pointing toward several pathways of design and participatory user consultation that could expand our understanding of digital well-being onto more politically dynamic terrain. Ultimately, the aim of this article is to reinvigorate the issue of user well-being as a nexus point of political concern, through which researchers from across multiple disciplines can study experiences of digital distress as symptomatic of wider social inequalities and (capitalist) relations of power. In doing so, we hope to open new avenues of critical analysis and opportunities for reflection on digital well-being, while lessening the unfair burden that individual users currently carry to fix wide-spread, and complex, contemporary sociotechnical harms.  

%% file: contents/05_well-being.tex
\section{What is well-being?}
\label{sec:well-being}

\subsection{Distinguishing between Health and Well-Being}

Health and well-being, while often combined, refer to two distinct aspects of human vitality. When thinking about computational systems that target health and well-being, therefore, it is important not to collapse the latter designation into the former. The World Health Organization (WHO) defines health as ``a state of complete physical, mental and social well-being and not merely the absence of disease or infirmity''~\cite{wilkinson_social_1998}. This is a prominent definition and forms the basis of international, and nation-specific, humanitarian health interventions. However, integrating bio-physical functioning (physical well-being) with assessments of mental health and relative judgements of social well-being is not clear cut. Disability Studies and the Critical Medical Humanities, for example, show how human health is always a discursive formation that mobilizes assumptions about normal and pathological bodily functioning~\cite{viney_critical_2015, puar_prognosis_2009}. These have been shown to often rely upon normative, racialized, and gendered conceptualizations of the human being as ideal types~\cite{bailey_misogynoir_2021, canguilhem_normal_1991}. Nevertheless, determining healthy physical and mental  human states can still be usefully, if not fully comprehensively, established through clinical trials and measurements. This measurability, as will be explored further below, distinguishes it from more general descriptions of well-being.

Computation could provide diagnostic support for practitioners through automated symptom detection~\cite{paparrizos2016screening, balagopalan2020cross} and medical test interpretation~\cite{kononenko2001machine}, while Machine Learning and Artificial Intelligence have been applied in several clinical ``rehabilitative, surgical, and predictive''~\cite{secinaro_role_2021} spheres. The field of digital health in general encompasses a range of technological interventions, such as clinical telepresence, healthcare applications, data management systems for hospitals, digitized devices for delivering pharmaceuticals (such as insulin pumps), patient self-monitoring tools, and robotic surgery, amongst innumerable others~\cite{lupton_digital_2015}. Machine learning is also becoming prevalent in detecting symptoms for mental illness~\cite{thieme_machine_2020}, and other digital techniques have been incorporated into psychological diagnostics. For example, browsing histories have been used to detect eating disorders~\cite{sadeh2020predicting}, and social media data has been drawn upon to detect mental health conditions, including depression and cases of suicidal ideation~\cite{de2016discovering}.

However, ~\citet{birk_for_2021} have cautioned against over-reliance on this type of `digital phenotyping', arguing that behavioural data can never be assumed to be a simple reflection of innate user traits. Rather, data traces emerge through complex processes of mediation~\cite{latour_technical_1994}. Here, the co-production of situated users and digital devices complicates any simple linearity between the extraction of behavioural data and actionable health insights. Beyond individual health support, and despite this, behavioral data has inspired the creation of automated tools in the service of public health, such as Google’s (now defunct) system for detecting seasonal flu outbreaks using temporal and spatial trends in the frequencies of search queries~\cite{ginsberg2009detecting}. More recently, similar approaches have been tested for early detection of Covid outbreaks~\cite{lampos2021tracking}. By being grounded in established medical categories and modes of  measurement, notwithstanding the perhaps problematic values and epistemological biases implicit within them, this broad landscape demonstrates that digital interventions in public, physical and mental health have been effective in several fields. 

Intervening in human well-being, in contrast, is less secure due to fundamental issues to do with its definition. Well-being is a relative concept in two related ways. First, the "well" in well-being mobilizes normative judgements about how humans should best live their lives. Second, the `being' in well-being requires an ontological definition of what the human being is. Determining what is good for the human, as such, is always an evaluation of what the human individual \textit{is} and \textit{ought to be}. Therefore, whereas biophysical functioning helps ground categories of health, definitions and measurements of well-being necessarily draw upon more tenuous and contingent codifications. Vague references to human nature often fill this gap and are almost always implicit in suggestions of activities that lead to human flourishing. However, the study of human nature is decidedly controversial ~\cite{dupre_human_2001}. While evolutionary psychology could theorise a model of human nature linked to biological functioning within a social context~\cite{roberts_communication_2011}, the jump to then associate these biological processes with human well-functioning raises deeply ethical issues. This is particularly clear when we recognize the way oppressive modes of social organisation are often justified upon the basis of supposedly `natural’ hierarchies of human groups. To give one example, patriarchal discourses extant in early 20th century UK pointed toward the `natural’ intellectual inferiority of women to argue against female suffrage~\cite{mayhall_creating_1995}, which we can still observe as common feature of contemporary misogyny~\cite{braidotti_posthuman_2013}. Elsewhere, colonial powers justified the exploitation of labour power and resources on an image of the natural `savagery’ of indigenous peoples~\cite{mathieu_dynamics_2018}, while the genetic classification of racial groups remains an operative element of racism today~\cite{roberts_fatal_2011}. As a result, we should be wary when certain styles of life are presented to us as `natural’ to human well-being as opposed to others. When we encounter such suggestions, we ought to question where these modes of human flourishing originate, who is expressing them, and what interests could be served through their inculcation.

\subsection{Categories and Measurements of Well-Being}

Despite these  difficulties of definition, well-being has broadly been categorized using three descriptions: 1) eudaimonic well-being, 2) hedonistic well-being, and 3) structural accounts of well-being. How one understands well-being determines how it can be measured. It is therefore important to fully understand the stakes of incorporating different notions of well-being into computational design processes.

Eudaimonic well-being involves determinations of what it means to \textit{live well} and finds its roots in Hellenic concerns with the good life~\cite{crisp_aristotle_2014}. In the contemporary philosophy of eudaimonia, well-being is presented as a process, where one strives to fulfil innate capacities in order to flourish. Lorraine Besser-Jones (2016)~\cite{besser-jones-eudamonia} describes this as `well-functioning'. Here, well-being is active and ongoing, as opposed to a nominal state, and explicitly carries a morally normative dimension. Second, hedonistic conceptualizations of well-being, as the psychologists Ryan and Deci (2008) state, ``focus on a specific outcome, namely the attainment of positive affect and an absence of pain''~\cite{ryan_living_2008}(p. 140). This version of well-being resonates with Utilitarian philosophy, originating in England in the mid to late eighteenth century. This system of thought is best expressed in Jeremy Bentham’s phrase that ``the measure of right and wrong'' can be found in what brings ``the greatest happiness of the greatest number''~\cite{harrison_bentham_1988}(p. 3). Third, structural accounts of well-being attempt to analyze experiences of well-being in relation to relative socio-material circumstances. Specifically, structural accounts draw attention to the way configurations of race, gender, disability, education, and class, amongst others, intersect with lived environments to contribute to well-being, both in the hedonistic and eudaimonic senses just mentioned~\cite{brown_responsible_2012}. Cultural geographers, Smith and Reid (2018), for example, advance an `intra-active' account of well-being, which acknowledges the importance of ``place, space and context'' in the experience of well-being~\cite{smith_which_2018-1}(p. 807) – as identified in factors such as income, type of accommodation, and local levels of pollution.

How one conceptualizes well-being determines what factors to incorporate in its measurement. For example, recent psychological accounts of eudaimonic well-being have attempted to establish the universal needs requisite for sufficient levels of human flourishing~\cite{haybron_pursuit_2008}. Martela and Sheldon have identified 45 different versions of eudaimonic well-being in the literature, using 63 different measures between them~\cite{martela_clarifying_2019}. ~\citet{costanza_quality_2007}, for example, draw upon Maslow’s hierarchical pyramid of human need~\cite{maslow_motivation_1954} to suggest that subsistence, reproduction, security, affection, understanding, participation, leisure, spirituality, creativity, identity, and freedom are the basic living factors necessary for quality of life. The authors suggest self-reporting methods to measure the extent to which individuals feel these factors are being met, which can then be aggregated to give a picture of well-being at wider population scales.

Measurements of Subjective Well-being (SWB) constitute attempts to study well-being from a hedonistic perspective. ~\citet{diener_advances_2018} classify SWB as ``a person’s cognitive and affective evaluations of his or her life'' (p. 63). SWB is assessed in terms of three main components: the presence of positive mood, the absence of negative mood, and life satisfaction~\cite{kahneman_personality_2003}. At a very basic level, measuring SWB involves participants answering questions about how they feel about their life and circumstances. Various surveys and scales have been developed that seek to assess this~\cite{diener_advances_2018}. For instance, the Satisfaction with Life scale~\cite{diener_satisfaction_1985} requires participants to respond to face-value statements such as ``I am satisfied with my life'' on a Likert scale of agreement. Self-report surveys also exist that assess how frequently and intensely subjects experience positive or negative emotions over a certain time-period, such as `happy,' `sad,' `angry,' or `joyful'~\cite{diener_new_2010}. In this framework, how happy one feels in their day-to-day lives, and how satisfied one is with the direction of their life in general, is the basic correlate of well-being.

Finally, as the well-being of the individual in structural accounts cannot be divorced from their lived circumstances, measuring the contextual supports that shape experiences of well-being are as important as subjective experiences themselves. This focuses our attention to the societal, political, and cultural factors that make life more or less difficult on a daily basis for individuals in different contexts. These wider factors can be usefully approached through the social determinants of health framework, which refers to the impact ``the conditions in which people are born, grow, live, work and age''~\cite{who_commission_on_social_determinants_of_health_closing_2008}(p. 1) has on well-being. This framework recognizes that the material circumstances that facilitate well-being are uneven, and often intersect with systemic differentiations of class~\cite{wilkinson_social_1998}, gender~\cite{raphael_social_2008}, race~\cite{yearby_structural_2020} and education~\cite{shankar_education_2013}, amongst others. This stands in contrast to the notion that well-being is somehow a personal achievement, deriving simply from individual thought, choice, and action alone. Structural accounts of well-being implicitly recognize the difficulty of creating universally comparable metrics of well-being that can account for basic differences in lived experience, which emerge from the vastly unequal distribution of wealth, land, and resources we currently find within structures of contemporary global capitalism. Questions we can ask here could include: How do metrics of income impact well-being? How do various living arrangements impact well-being? How does relative access to health care impact well-being? Teasing out the complex implications of this relational view of well-being in HCI will be explored in more depth in the closing sections of this article, and will be presented as a prerequisite for socially conscious design.

\subsection{Digital Well-Being}

Within this contested terrain, what does it mean, then, to design for user well-being? What particular versions of well-being do we commonly encounter in HCI? What is being left out? Although notable studies have sought to explore metrics of eudaimonic well-being in fields of User Experience~\cite{mekler_momentary_2016}, hedonic perspectives that centre utility and user pleasure are historically influential in the HCI literature~\cite{diefenbach_hedonic_2014}. Subjective Well-Being (SWB) scores are often used to measure the impact of particular technologies on user well-being~\cite{desmet_positive_2013}. This is especially prevalent in research on social media, which correlates self-reported measures of SWB with language usage patterns in social media posts~\cite{chen_building_2017}, time spent on social media~\cite{kross_facebook_2013}, or different types of `active' use and `passive' use~\cite{verduyn_social_2017}. However, some scholars have pointed out the limitations of engaging with such metrics in the study of well-being. \citet{smith_which_2018-1}, for example, argue that SWB measures assume a knowing subject able to report on their feelings of happiness that may not stand up to scrutiny. For example, psychoanalytic perspectives would question the ability of fragmented selves to reliably access their `true' feelings in response to such questioning~\cite{parker_psy-complex_2018}. Moreover, while it would be common-sense to assume that happiness is universally desirable, \citet{atkinson_toxic_2020} argues that SWB measures presume the existence of an `hyper-individualized' subject that is in fact tied to a particular historical model of liberal individualism, and a correspondingly atomistic conceptualization of society made up of discretized social networks. For example, \citet{christopher_situating_1999}  highlights how placing the ``onus of well-being on the individual'' (p. 143) further mobilizes a Western-centric conceptualization of the self and resonate models of personal responsibilization. This stands in contrast to more collectivist traditions, which factor in duty, obligation to others, and social harmony as a key aspect of well-being~\cite{markus_culture_1991}. Despite its ostensible neutrality, therefore, the type of happiness that SWB measures is not a value free concept, and is, in fact, inseparable from the temporal and cultural location of its study and expression. This again underscores the normative issues at stake in how we choose to conceptualize and measure well-being in the design process. 

Recent work in HCI embraces the potential of technological systems to intervene in and facilitate human flourishing. Some frameworks explore the links between technological design and positive psychology, explicitly seeking to design for user pleasure, personal meaning, and virtue~\cite{villani_integrating_2016}; other approaches, such as positive computing, examine how empathy, mindfulness and compassion can be incorporated into the design process~\cite{calvo_positive_2014}; while work on positive technologies looks to improve affective quality, actualization, and connectedness for users~\cite{riva_positive_2012}. As has been demonstrated above, these are all unavoidably normative prospects, as the emphasis on `positive’ in such frameworks makes clear. However, in all these models the implications of this normativity are made explicit. Because such work clearly lays out the theoretical, empirical, and psychological stakes of intervening in positive human functioning, readers, practitioners, designers, and users can all assess the extent to which such interventions align with their own beliefs, values, and needs -- at least in theory. This type of transparency could factor into attempts to create more socially equitable computational systems by leaving room for clear critical interpretation, contestation, and assessment. Within current apparatuses of capitalist production, whether platforms and designers can be incentivised, or even \textit{made}, to be open to such consultation (if at all) raises larger questions that we will return to later on.

Nevertheless, such approaches resonate with other critical design frameworks such as Philip Agre’s~\cite{agre_computation_1997} mode of Critical Technical Practice (CTP), which incorporates reflexive philosophical questioning into the planning phase, and forms of Value-Sensitive Design (VSD), which likewise brings to light the relativistic value judgements implicit in technical systems~\cite{friedman_value_2019}. VSD constitutes a varied and rich intellectual field that we do not have the space to fully explore here, yet a core insight that we build upon is the notion that technologies, far from being neutral, embody moral intentionalities by facilitating some forms of ideal usage, and restricting others~\cite{flanagan_embodying_2008}. VSD highlights the problems that arise when designers assume user universality, and propose one-size-fits all ideal engagements from situated, culturally specific, standpoints. In the case of digital well-being, this would involve assuming that well-being constitutes the same thing for all users, in all locations, at all times -- which the previous sections have shown to be impossible. To manage the potential exclusionary designs that could arise from maintaining such blind spots, VSD endorses a range of creative participatory techniques that aim to incorporate a variety of subject positions, voices, affects, and senses into the design process~\cite{friedman_value_2013}. The potential, and desirability, of participatory approaches to make room for multitudinous perspectives to be both represented in and facilitated by interactive technologies is something we will critically explore in more depth in the discussion section.

\subsection{Good for whom?}

In the words of Ng et al. (2003) ~\cite{ng_search_2003}: ``understandings of well-being are clusters of cultural assumptions and values [...] they necessarily rely on moral visions that are culturally embedded and frequently culture specific'' (p.323).  Despite this recognition, expansion patterns of contemporary digital platforms raise the concern that certain culturally specific norms of well-being are becoming globally expressed at the expense of others. Digital technologies frequently emerge in the Global North, saturate these markets, and then seek to win over the ``Next Billion Users'' -- a term which major tech companies like Google have used for the user populations in the so-called Global South~\cite{google-next-billion}.  Westernized notions of well-being as an individualized, personal responsibility (as expressed in the Subjective Well-Being frameworks) might thus flow to these new markets encoded in technological designs. Yet, as illuminated by Arora~\cite{arora2019next}, this expansion may not represent the values and needs of users in other regions of the globe. Arora shows how Westernized assumptions about the technological needs of users in the Global South are often inaccurate, if imagined at all. Similarly, \citet{sambasivan2021re}, using India as a case study, reveals how value judgements implicit in discourses and designs of algorithmic fairness, which have been developed within liberal ethical-legal frameworks, often fail in regions outside its initial design context. This is due to vastly differing conceptions of user group identity, membership and conceptions of justice that exist elsewhere around the globe.

This demonstrates how the value systems that are embedded in technological designs may not be relevant, or potentially even harmful, to users living differently to the ways assumed by situated designers. This point is emphasised by the work of scholars that reveals how centering different philosophical value systems can actually lead to very different actionable technological design proposals. ~\citet{wong2012dao}, for example, has proposed a Confucian approach to digital ethics, focusing on harmony and social roles, while recent Indigenous AI protocols seek to incorporate diverse Indigenous knowledge systems into building responsible AI~\cite{lewis2020indigenous}. Rigidly sticking to one conception of well-being in the design stage is to impose a set of norms, even in sincere developments such as globally accessible internet infrastructure and discourses of algorithmic fairness. To build on a core argument this paper has been making so far, which we will return to in more depth later sections, we propose that these norms should at least be made explicit in the design and release of technologies aiming toward digital well-being, rather than simply be taken for granted as universally applicable to all peoples at all times. 

We have shown how well-being is always a culturally relative proposition tied to the time and place of its articulation~\cite{christopher_situating_1999}. This makes designing appropriate, and globally applicable, computational interventions in well-being problematic, if possible at all. When we reflect upon the discourses and designs of digital well-being currently disseminated today, it is crucial to be mindful of which well-being is being expressed, who is expressing it, and for what specific purposes. The following section develops this ethos of critical transparency in order to uncover the normative stakes and limitations of conceptualizing and measuring well-being through one readily available present-day marker: user engagement metrics. This type of analysis serves a dual function. First, it will provide further conceptual clarity into how existing HCI metrics incorporate measures of digital well-being. Second, it will demonstrate why we should practice caution when doing so. 

%% file: contents/10_engagement.tex
\section{User Engagement as a Proxy for Well-Being?}

As discussed in Sec.~\ref{sec:introduction} and~\ref{sec:well-being}, recent approaches to digital well-being measurement focus on dimensions of time as well as user connection and disconnection. To ground our analysis, we generalize these dimensions under the umbrella of \emph{user engagement} -- a platform evaluation framework for quantifying the quality of user experience and time spent on a platform. This section examines a state-of-the-art user engagement measurement framework, before contrasting it with a commonly-used framework for digital well-being that focuses on technology addiction. A formal comparison between some of the established measurement methodologies for both engagement and user well-being will help us exemplify the issues pertinent to using the former as a proxy for the latter, as well as informing a discussion of alternative approaches.

\subsection{Metrics of User Engagement}

Metrics of user engagement quantify the "quality of the user experience that emphasizes the positive aspect of the interaction with an online service and, in particular, the phenomena associated
with wanting to use that service longer and more frequently"~\cite{hong2019tutorial}. The measurement approach is motivated by the fact that "people remember enjoyable, useful, engaging experiences and
want to repeat them"~\cite{attfield2011towards}. While engagement encompasses at least eight different dimensions~\cite{attfield2011towards}, the dimension of \emph{endurability} underlies many metrics, as it can be quantified from online behavioral logs~\cite{hong-lalmas-kdd, hong-lalmas-kdd-materials}. These metrics are used across types of platforms, including search, news, recommendation, streaming, gaming, e-commerce, and social media. This section briefly overviews engagement metrics as presented by~\citet{hong-lalmas-kdd}; we refer the interested reader to the full tutorial~\cite{hong-lalmas-kdd-materials}.

Platforms often employ suites of metrics, as typical user engagement patterns differ between application domains and different metrics will be more or less accurate in quantifying engagement. For instance, users of search platforms typically return to a platform frequently but spend relatively little time during each interaction, while social media users might visit the site frequently and stay long. Even within a platform, different system elements, functionalities, or modes of interaction might trigger different engagement patterns. Despite the fact that multiple metrics are needed to capture interactions, most metrics aim at quantifying the endurability of user engagement during individual sessions (intra-session metrics) and loyalty across sessions (inter-session metrics).

\emph{Intra-session metrics} aim at capturing user activity within a single interaction session. They can be divided into three categories. The first set of metrics quantify involvement (this category contains metrics capturing more passive engagement modes such as dwell time or play time). The second set of metrics quantifies interaction (with more active engagement measures such as click-through rate, number of likes, or number of skips). Finally, metrics quantifying the most active platform participation modes capture user contribution (number of posts, number of comments, uploading, etc.). \emph{Inter-session metrics} go beyond a single platform usage session and aim at capturing longer-term engagement patterns and loyalty to a platform (frequency of platform visits, total usage time across sessions per month, average length of disengagement periods between sessions, etc.).

Longer time spent on a platform will not necessarily imply higher engagement. Some metrics of involvement might increase (for instance, dwell time), but time spent does not necessarily lead to more active types of engagement. Engagement metrics are mere proxies for user engagement, which have been shown to work reasonably well in practice to drive product development and satisfy business needs. However, many questions and concerns arise should engagement metrics be used as proxies for well-being: low or high engagement metrics cannot unambiguously be attributed to high or low levels of well-being, as the remainder of this section will show.

\subsection{Case Study: User Engagement vs Digital Addiction}

To illustrate core problems with using engagement metrics as proxies for well-being, we analyze engagement vis \`a vis digital addiction,\footnote{Whether conditions such as internet or gaming addiction should be recognized as clinical disorders is 
subject of active research and debate.} a concept commonly used as tool for operationalizing digital well-being. The Digital Addiction Questionnaire (adapted from a questionnaire used for gambling addiction and a version of which has been developed for a variety of digital scenarios including internet use~\cite{young1999internet}) has been used as a ‘framework’ for quantifying problematic uses of technology. Dimensions of the questionnaire aim at capturing different demonstrations of pathological use, and a positive answer to any five of the eight dimensions is required for a clinically valid diagnosis~\cite{young2009internet}. 

\begin{table*}
  \small
  \begin{center}
  \caption{Pathological engagement criteria vs patterns in engagement metrics in case a criterion is satisfied.}
 \label{tab:case_study}
    \begin{tabular}{p{0.45\textwidth} p{0.45\textwidth}}
    \toprule
      \textbf{Pathological engagement criteria 
(Internet Addiction Test, Young 1998~\cite{young1999internet})}  & \textbf{Engagement metrics patterns in case criterion satisfied} \\ 
      \midrule
      
      C1. Do you feel preoccupied with the 
Internet (think about previous on-line activity 
or anticipate next on-line session)?  & Cannot capture (offline).  \\
\midrule

      C2. Do you feel the need to use the Internet 
with  increasing  amounts  of  time  in  order  to 
achieve satisfaction? & Increase in intra-session metrics; stagnating or decreasing value of satisfaction 
metrics.  \\
\midrule

     C3. Have you repeatedly made 
unsuccessful  efforts  to  control,  cutback,  or 
stop Internet use? & Can result in patterns such as: a period of low values of inter-session metrics 
(long  non-engaged  times  in  between  sessions)  followed  by a  period  of  high 
intra-session and/or inter-session metrics (relapse after unsuccessful cutback 
attempt).\\
\midrule

     C4. Do you feel restless, moody, depressed, 
or  irritable  when  attempting  to  cut  down  or 
stop Internet use? & Cannot capture (offline). \\
\midrule

     C5. Do you stay on-line longer than 
originally intended? & Cannot capture (in most cases a system would not be aware of the intended 
engagement duration). \newline %\\[0.5cm]

 Platforms  have  developed  predictors  of  user  engagement,  for  instance,  to 
increase the effectiveness of A\textbackslash B testing\text{~\cite{drutsa2015future}}. Even though such techniques could 
potentially  be  used  to  quantify  this  criterion,  engagement  prediction  errors might  be  impossible  to distinguish  from  a problematic  (longer  than  intended) 
platform use. \\
\midrule

    C6. Have you jeopardized or risked the loss 
of a significant relationship, job, educational, 
or career opportunity because of the 
Internet?  & Cannot capture (offline).\\ 
\midrule

    C7.  Have  you  lied  to  family  members,  a 
therapist, or others to conceal the extent of 
involvement with the Internet? & Cannot capture (offline).\\
\midrule

    C8.  Do  you  use  the  Internet  as  a  way  of 
escaping  from  problems  or  of  relieving  a 
dysphoric mood (e.g., feelings of 
helplessness, guilt, anxiety, depression)? & Cannot capture (offline). \newline
 
 Music streaming services suggest that they may be able to predict user 
emotions; part of this criterion could be captured through enhanced 
engagement modeling on such platforms. \\
    \bottomrule
    
    \end{tabular}
  \end{center}
\end{table*}

Table~\ref{tab:case_study} presents a mapping from the criteria of digital addiction to anticipated patterns in engagement metrics when a criterion is satisfied. There are only two dimensions of pathological engagement that are likely to incur observable changes in engagement metrics patterns: the need to increase the platform usage time to achieve satisfaction and repeated, unsuccessful efforts to control, cutback, or stop a platform use. These dimensions indeed relate to the time spent on the platform. Still, they relate in a non-linear manner. We expect to see increases in metric values over time, or complex engagement and disengagement patterns.

This exercise reveals that most dimensions of digital addiction cannot be captured by behaviorist metrics of on-platform behavior as they relate to aspects of a person’s life that are typically not accessible to the platform. Indeed, with few exceptions, information about a person’s lost personal and professional opportunities, or thoughts and emotional states when not using a platform, cannot be directly observed through behaviourist platform measurements. Even if information about a person’s relationships, jobs, or emotions could be automatically inferred by some online providers (e.g., from search queries), the inferences will be imperfect. This inaccuracy will in particular impact criteria that compare user behavior with user intentions, such as criterion 5 in Table 1. Here, if there is an observed discrepancy between the predicted and actual duration of user engagement, it might be impossible to attribute it unambiguously to an algorithmic error or a problematic engagement pattern.

Reversely, it can be problematic to use patterns in engagement metrics to infer the state of user well-being, as similar patterns of engagement could be triggered by other factors. For instance, erratic disengagement patterns might be caused by long-term travels and a resulting lack of a constant and reliable internet connection.

\subsection{Issues with Engagement as a Proxy for Well-Being}

The case study above has exemplified the limitations of approximating digital well-being (digital addiction) using engagement metrics. We now generalize these observations, pinpointing two key issues: A minimum viable well-being measurement will need to account for the intertwining effects of user intents and the platform context.

Classical taxonomies of user intents in search~\cite{broder2002taxonomy} generally distinguish three intent types: navigational (when a user has a specific website in mind), informational (when a user wants to gain new information about a topic or seeks advice), and resource queries (locating files, videos, or online tools). Search results that satisfy user intent are considered relevant, and the relevance signals are then used to evaluate the system using standard performance metrics. Intents in recommender systems are modeled differently in various application areas. For instance, music streaming services differentiate between Leaning In, Active, Occupied, and Leaning Back intents for different modes of listening (for activities like: music discovery, dancing, workout/studying, or sleeping, respectively)~\cite{hong-lalmas-kdd-materials}.

Well-being as care for the self can be assumed to be a reasonable motivation for each of these engagements. In navigational searches, a user might look for sites that provide information that could enable them to `live well'. Precisely what this information constitutes will be particular to the individual involved. For example, one user may seek out the nearest yoga studio, another may seek out the nearest bar. Both of these search terms, and actual practices, cannot be assumed to be better or worse for user well-being without mobilizing explicit value judgements as to what well-being constitutes for the user in question. In informational searches, moreover, a user may seek for more general advice about particular practices that could ameliorate their own well-being, while this could be followed by resources searches looking for particular facilitating objects toward this end.

Yet in some specific cases, when a user purposefully searches for information intended to facilitate self-harm for example, an `effective' system satisfying user intent could lead to outcomes that may seem in contradistinction to the imagined well-being of users. This is not to moralize and medicalize self-harm, but is to rather highlight the gray areas involved in configuring what constitutes an appropriate digital well-being intervention. Such interventions are already in place. For instance, in some of the more explicit cases of self-harm intents, search engines `inject' interventions to allow users to find help resources instead (queries related to suicide, for instance, trigger additional results related to suicide prevention in search engines such as Google or Bing). What is more, a user might end up exposed to potentially damaging results even without a clearly self-detrimental intent. To give one example, infrequent search queries could be hijacked to spread misinformation or propaganda, a problem termed `data voids'~\cite{golebiewski2018data}.

Finally, some studies have found that personally distressing digital engagement may be a compensatory strategy for coping with pre-existing life problems. For instance, Kardefelt-Winther found evidence of escapist motivations behind excessive gaming in people with high stress levels and low self-esteem~\cite{kardefelt2014moderating}. These findings raise the question of whether algorithmically reducing engagements can be an effective intervention without tools to address the deeper psychosocial problems causing problematic engagements in the first place. This is a key point we will return to in the discussion section.

The discussed problems exemplify two of the core issues of using engagement metrics as a proxy for well-being: 
\begin{itemize}
    \item     A similar surface pattern of engagement on different types of platforms might be a quantification of both increase and decrease in well-being.  

    \item     A similar surface pattern of engagement on the same platform by different users might be a manifestation of both well-being increasing and decreasing behaviours.  
\end{itemize}

Differentiating between `positive' and `negative' well-being engagements on a platform will not be possible without (i) normative accounts of what well-being constitutes (ii) modelling of user intents, and (iii) personalization (modeling of user background and persona), at the very least. Yet, to achieve an accuracy level necessary for technical well-being modelling, a platform may have to resort to invasive designs and data collection practices. The next section discusses the limited appeal of these solutions and how they can be best avoided through alternative interventions. 

%% file: contents/15_discussion.tex
\section{Discussion}

\subsection{Potential Designs}

\subsubsection{Expanded Modeling and Invasive Designs}
Our analysis has demonstrated that nuanced modeling of user well-being and its inclusion as a systemic objective would necessitate expanded user modeling and personalization. Models would need to account for offline aspects of user life and deepen the understanding of user engagement intents and their targets. Such modeling might be impossible without invasive data collection practices: collecting intimately personal information via surveys, modalities other than click or time-spent information, or through automated inference from existing signals. Allowing platforms to collect this gamut of sensitive information invites a reflection on the downstream negative consequences for privacy, user autonomy, and the power balance between users and platforms. To be clear: this invasive data collection is not a `solution' to the complexity of digital well-being that we endorse. Nevertheless, whether the benefit of automating well-being measurement could balance these (undesirable) consequences remains an open question that designers committed to the task must face. As we are in no way suggesting that designers and platforms \textit{should} expand these more invasive data collection protocols, we will now discuss alternatives.

\subsubsection{Managing Negative Externalities and Reducing Harm}
If expanded modeling is not a justified means, managing negative externalities of platform engagement through harm reduction frameworks could be one alternative. Recognizing, mitigating, and preventing damaging user effects can in many cases be seen as an intervention for user well-being. An example of such an intervention is limiting user exposure to problematic online content. In this sense, ~\citet{singhbuilding} seek to increase user well-being through reinforcement learning solutions aimed at minimizing user exposure to violent videos in automated video recommendation sequences. Similarly, \citet{ribeiro2020auditing} audit Youtube video sequences that could lead to political radicalization. \citet{das2020fast} propose solutions to enhance the well-being of content moderators and crowdsourcing workers through blurring disturbing content in photos in annotation tasks. A different approach is taken by positive computing frameworks, which, as introduced earlier, target eudaimonic user well-being in order to create the conditions for joy, self-expression and flourishing to emerge.

Such interventions, whether preemptive or reactive, are commonly viewed as harm reduction techniques, which aim at reducing the negative effects of problematic technology use, without necessarily seeking to extinguish such usage together. Although this approach is sometimes met with resistance in health contexts, particularly around drug treatments~\cite{hawk_harm_2017}, it offers a less morally punitive framework of care – one better able to recognize and respond to the intersecting social, political, cultural, and environmental pressures that may have led to problematic usage of technology to emerge in the first instance. In this vein,~\citet{swanton_problematic_2021} have developed a stakeholder framework to minimize harmful use of emerging technologies, which ~\citet{kuss_risk_2021} suggests highlights corporate responsibility as one key requirement for the future safeguarding of user well-being.

\subsubsection{Value-Sensitive and Participatory Designs}
We have shown how any appeal to well-being is implicitly normative. Therefore, knowingly or not, designs that do not attend to the various structural issues, cultural norms, and individual differences that condition diverse experiences of user well-being are engaged in a normative task. Designs which focus solely on individual user engagements, as such, are perpetuating particular accounts of well-being, which may be irrelevant, or even harmful, to different users from different backgrounds in different regions. Reversely, work that is attentive to the various impacts that unequal socio-material circumstances have on different users, as well as the global variety and philosophical diversity of well-being, could produce more inclusive and sustainable technologies.

The conceptual, empirical, and technical investigations of the value-sensitive design process~\cite{friedman_value_2019} can help creators elicit value inputs from direct and indirect system stakeholders, understand their goals, and adapt technological affordances to offer users maximum scope for flexible interactions. Friedman~\cite{friedman1996value}, for example, describes how lower system complexity can enable users to fluidly interpret technical objects, according to their own requirements.  In the context of digital well-being, we could envisage a system transparent in its physiological, cognitive, and emotional risks toward users, and one that enables users to act on their personal engagement intents rather than algorithmically inferring and enforcing them. Alongside this, technologies designed for well-being could be explicit in their determinations of what counts as `living well' with the technical object. Such clarity could, in theory, enable users to determine whether or not the design aligns with their own values. 

Participatory approaches are a further alternative to invasive data collection practices, incorporating the insight of users in the design process. This is to say that users themselves are probably best positioned to decide what ‘well’ means for them considering their distinct cultural and socio-material circumstances. Individuals might be best positioned to decide which types of digital engagement are most and least conducive to their personal well-being as a result, and what modes of measurement could potentially capture their own interpretations of well-being. Co-design can fulfill two functions. First, it might help open the digital well-being design process by incorporating the perspectives of different system stakeholders as well as the affected communities. Second, since well-being does not have a stable interpretation, it might help designers navigate its individually fluctuating meanings. Developing participatory design practices for well-being modeling, either through community consultations for high-level system designs, or within-system input facilitation from individuals for personalized interventions, constitutes a potentially productive direction for future research.

However, these types of design interventions, which could offer users the chance to become designers themselves, may place an unwarranted, and unasked for, `safeguarding' burden on individuals. It could be that participatory designs of well-being could end up repeating the same focus on technological interventionism as a `fix' to decreased user well-being, while ignoring the structural factors that we know are hugely significant. Since many well-being harms are inflicted by systemic infrastructures that perpetuate inequity, participatory approaches could be complemented by offering tools for critique that could draw attention to these factors. Here, users could be encouraged to explore the risks and the mechanisms that create feelings of personal technological distress, specifically within the context of digital capitalism and the extractive operations of the attention economy. Section 4.3 will explore how this type of grassroots critical orientation could open political pressure points, with the potential to lead to more systemic change on a political and societal level. This is to not endorse a naive belief that better technologies inevitably lead to the creation of a better world. Rather, we wish to focus on how the \textit{process} of considering what a better world could look like, and how technology functions within these visions, could make room for new ideas and new ways of life to emerge - however slim this hope remains.

\subsubsection{Fairness in Design}
As described earlier, structural accounts of well-being consider people’s socio-material circumstances and center the relational and community aspects of well-being. In this light, recent work on algorithmic fairness, in particular work that focuses on improving outcomes for an under-served group of users, could be seen as sensitive to the issues raised within structural accounts of well-being. Moreover, one of the pressing questions in algorithmic fairness concerns the acceptance of fairness interventions by stakeholders who might be inconvenienced by the outcomes. The connection we have drawn between fairness and structural well-being, as well as the cultural roots of different conceptions of well-being, suggest that the issue is perhaps culturally-dependent. In cultures with dominant structural conceptions of well-being, acceptance of fairness interventions might come naturally. In cultures with dominant utilitarian and hedonistic conceptions, acceptance might require a deeper alignment of individual interests of different groups. The literature on inclusive (or equity-focused) teaching could provide a framework to recognize and champion these differences~\cite{baglieri2004normalizing}. For example, Molbaek~\cite{molbaek2018inclusive} highlights how widening what constitutes educational success beyond single-dimensional performance foci actually creates new opportunities for student growth and achievement. Similarly, widening the scope of digital well-being beyond user engagements offers designers a chance to rethink what user well-being actually is, potentially opening space for a more inclusive understandings and experiences of flourishing to emerge.

\subsection{Non-design}
While positive design, VSD, harm reduction techniques, and inclusive design practice are applicable responses to the divergent well-being impacts technology has on users, a more radical implication, as expressed by ~\citet{baumer2011implication}, could be to simply not design these potentially problematic technologies in the first place. While non-design may appear counter-intuitive to the aims of HCI, it should nevertheless be a legitimate option for practitioners. If the risks of problematic engagement with technologies appear too high in the design stage, then perhaps there is a duty to not continue the project. Likewise, if certain technologies seem to be having continued deleterious impacts on users, there could be calls to withdraw such technologies from the market altogether. These principles of non-design run in contradistinction to the principles of capitalist product research, design and marketisation. They may appear outlandish as a result. However, perhaps this rupture in the norm surfaces an opportunity to reflect upon the very nature of capitalist technology development in the present age. What level of user care is reasonably expected in capitalism? What is driving the design and release of these potentially harmful technologies? Who is responsible for user well-being? These are significant questions that require more time and a greater scope than the aims of this paper allow here. Yet, situating these design strategies and assessing their validity within wider historical and sociopolitical contexts is vital if we are to fully approach the issue of user well-being with the clear eye and analytic comprehensiveness needed.

\subsection{Re-politicizing Digital Well-Being}

As the structural accounts of well-being introduced above show us, well-being is not the sole outcome of personal behaviors. Even if we could accurately account for user intent in different platform contexts, distilling user well-being down to the behaviors that are measurable through engagement is insufficient. Such an approach fails to assess how circumstantial factors impact user well-being prior to users’ engagement with various platforms. Problematic relationships with technology, which may or may not be termed addiction, do not emerge within a vacuum. ~\citet{chen_smartphone_2020}, for example, shows how social factors such as self-esteem impact detrimental use of social mobile applications, while ~\citet{mustafa_internet_2020} link internet overuse with existent levels of high family stress. In order to advance a more socially conscious notion of well-being in the design process, we could seek to pay greater attention to these circumstantial user experiences in our measurements of well-being. Doing so would push our analysis of digital well-being further than is currently imagined. Using the first of these studies as a jumping off point, for example, we may choose to explore how personal experiences of intersecting racist, misogynistic, classist, transphobic, or homophobic discrimination impacts perceived levels of self-esteem for different users. In the second case, we could investigate how household income, housing quality or access to childcare impact the levels of family stress that shape different types of technological take-up.

Building on the insight that such experiences of socio-economic inequity could, in some instances, be considered traumatic, the literature on trauma-informed design~\cite{dietkus} offers pathways to design technologies sensitive to the differentiated experiences of users. We might consider engaging in trauma-informed approaches developed in clinical settings to do so ~\cite{reeves2015synthesis}. For example, the Substance Abuse and Mental Health Services Administration framework~\cite{samhsa} defines six principles for working with patients with traumatic experiences. These principles include the fostering of safety, trustworthiness and transparency, peer support, collaboration and mutuality, empowerment, voice and choice, as well as sensitivity to cultural, historical, and gender issues. Trauma-informed practice, however, also teaches us that care is needed when designing for well-being, which may solicit memories of sensitive personal experiences resulting from inequity and oppression: exclusionary technological designs could create possibilities of re-traumatization. Indeed, Hirsch~\cite{hirsch2020practicing} has argued that we need novel trauma-informed research practices that could serve as a therapeutic process for users with traumatic experiences.

Drawing attention to these multitudinous contextual factors is necessarily a political endeavor, and shows how conscious theoretical and methodological consideration can center social justice issues at every stage of sociotechnical design and analysis. Such an approach resonates with recent cutting-edge medical research into the \textit{exposome}, which holistically explores the overlapping, and wide-ranging, ``chemical, social, psychological, ecological, historical, political, and biological elements''~\cite{marya_inflamed_2021}(p. 32) that impact health and illness. This type of critical approach currently re-emerging in medicine also chimes with the work of cultural theorists such as ~\citet{ngai_ugly_2007} and ~\citet{cvetkovich_depression_2012}. Such authors argue that negative human emotions, as well as mental health issues such as depression, can be used to highlight the effects of power as it pressures different individuals in different ways in the same societal environments. The feelings of anxiety, envy, and depression we often observe on social media platforms~\cite{tandoc_is_2021}, for example, ought not simply be viewed as the result of individual psychic deficiencies, as contemporary (neoliberalized) accounts of individualised digital well-being may have us believe. Rather, these negative feelings could alternatively be used as entry points to map and investigate the personal costs of living in social regimes, and the sociotechnical infrastructures which constitute them, that are hostile to certain ways of life.

Here, the identification of relative depreciations of well-being on platforms can provide a vital opportunity to discuss the psychic and intersecting impacts that racist oppression, misogyny, class subjugation, exploitative labour arrangements, homophobia, transphobia, and ablism may have on individual users. This would allow us to examine how these factors entwine with existing cultural milieus and various sociotechnical affordances. Doing so would treat the psychological issues manifest through specified forms of user engagement as social issues, as opposed to simply being treated as personal behaviors atomized from the environing conditions that support and sustain them. This relational viewpoint configures depreciations in well-being in digital environments as symptomatic of operative inequalities in the world, such as those emergent in contemporary structures of neoliberal capitalism~\cite{rose_our_2018}. This, as a result, grants a political force to the exploration of digital well-being that may be lacking in other accounts.

We have shown how measuring these exigent external factors is beyond the capability of user engagement metrics. However, failing to recognize the limits of user engagements as a proxy for well-being, and without advocating the need for them to be supplemented by other structural well-being metrics, has the effect of rendering these salient social factors as unimportant. This is not only empirically limiting, but also functions as a prescriptive gesture that produces constraining political thresholds that taint how we understand the issue of digital well-being at large. Ignoring the social, cultural, and economic inequalities that produce differences in user well-being is as much to say that these inequalities do not matter. As a result, the political necessity of ameliorating these inequalities is shut down. This enables existing power structures to remain in place unchecked, and the opportunity to link digital well-being with wider social justice issues is lost. HCI practitioners could draw upon the rich intellectual tradition of the social sciences and humanities to make these connections, or incorporate such modes of analysis directly into their work. Whichever approach is adopted, it is crucial for HCI experts to recognize that the way we conceptualize and measure digital well-being has serious normative political implications. Well-being should be handled with care as a result. More than this, however, this article has shown how modes of critique specifically drawn from Critical Theory and Cultural Studies can imbue this analytic of care with a distinctly political edge. Digital well-being represents an opportunity to discuss wider structural inequalities, not just an individual problem to be fixed through technical interventions. In this way, digital well-being can open up new (politically engaged) orientations for HCI researchers to explore, examine and expand in the future.

%% file: contents/20_conclusions.tex
\section{Conclusions}

This paper has highlighted the empirical and political limitations of adopting a purely behaviourist account of well-being in our attempts to design benign interactive computational systems. We have shown that the social, cultural, environmental, and material circumstances that condition well-being are often missing from existing accounts of digital well-being, which is often understood solely in terms of measurable user engagements. We have argued that this impoverishes our understanding of the issue at large. Returning to an established argument in HCI ~\cite{bannon_human_1995}, we instead wish to reassert that users have lives that shape their well-being engagements with different technological platforms prior to their arrival upon them.

Platforms are sociotechnical systems that are cradled in a web of historically constituted social, political, and economic dynamics. Impacts on user well-being, positive or negative, as a result, cannot be separated from these dynamics. Therefore, our attempts to understand and model for user well-being ought to be fully cognizant of these shifting elements as they are expressed in digital spheres. This unavoidably draws our attention to the various structural factors, which this article has explored in detail throughout, that we know play a crucial role in subjective experiences of well-being. Rather than study digital-well-being in isolation from these other contributing factors, which include personal differentiations of class, race, income, housing, gender, or nationality for example, future studies could instead foreground these factors in their analysis. This would further reveal that user well-being is more than the behavioural sum of its parts.

Once our attention is drawn to the structural factors that condition healthy activity, both `on' and `offline', the issue of user well-being is revealed to be something far more complex than simply managing time well spent on platforms. Through this new lens, designing systems that can nudge users toward healthier usage, in the expectation that this could broadly solve depreciated experiences of well-being on certain platforms, is a lot like expecting a band-aid to heal a broken leg. It may help by orienting our diagnostics to some degree, but alone it can in no way address the true scale of the problem area.

For HCI practitioners designing platforms with the well-being interests of users in mind, this framework makes previously implicit socio-political factors explicit, generating new sets of questions to engage with as a result. How can we design evaluations and optimization objectives that model and measure well-being for a particular person, accounting for their socio-material circumstances? To avoid imposing unwarranted norms, how can we co-design well-being approaches with communities and users? Should designers even demand this of users? Or adjust their expectations and view their role as managing inevitable negative externalities, rather than eradicate harm completely? Finally, is it appropriate to design for well-being at all, or will any effort be necessarily a ``computationally tractable transformation of a problem''?~\cite{baumer2011implication} We may wonder whether commercial platforms, as economic enterprises, are simply constituent parts of an integrated system of inequality; perhaps we should pay greater heed to the real limits of living well in extractive regimes of datafied life and exploitative modes of capitalist accumulation -- rather than simply trying to design our way out of them. These questions may not be comfortable, yet critically engaging with them could help us all understand, if not fully ameliorate, the full social parameters and political complexity of digital well-being as expressed in present-day sociotechnical networks.